\documentclass[journal]{IEEEtran}
\usepackage{amsmath,amsfonts,amssymb}
\usepackage{algorithmic}
\usepackage{algorithm}
\usepackage{array}
\usepackage{acronym}
\usepackage{textcomp}
\usepackage{stfloats}
\usepackage{xcolor}
\usepackage{url}
\usepackage{multirow}
\usepackage{hhline}
\usepackage{graphicx}
\graphicspath{{Figures/}}
\usepackage{cite}
\usepackage{balance}
\usepackage{orcidlink}

\usepackage{subcaption}
\def\BibTeX{{\rm B\kern-.05em{\sc i\kern-.025em b}\kern-.08em
T\kern-.1667em\lower.7ex\hbox{E}\kern-.125emX}}

\input{AcronymsListFinal}
\acresetall

\usepackage{hyperref}
\hypersetup{colorlinks=true, linkcolor=blue, citecolor=blue, urlcolor=blue}

\usepackage{fancyhdr}
\begin{document}
\title{Joint User Association and Pilot Assignment via Phase-Shifted Pilots for Scalable Cell-Free mMIMO}
\author{Mümtaz Özlen~\orcidlink{0009-0000-3804-9094}, Ahmet Sacid Sümer~\orcidlink{0000-0001-8866-8520}, \IEEEmembership{Graduate Student Member, IEEE}, Ahmed Naeem~\orcidlink{0000-0002-1534-5883}, and Hüseyin Arslan~\orcidlink{0000-0001-9474-7372},~\IEEEmembership{Fellow,~IEEE}

\thanks{Mümtaz Özlen, Ahmet Sacid Sümer, and Hüseyin Arslan are with the Department of Electrical and Electronics
Engineering, Istanbul Medipol University, Istanbul, 34810, Turkey (e-mail: mumtaz.ozlen@medipol.edu.tr, ahmet.sumer@std.medipol.edu.tr, huseyinarslan@medipol.edu.tr).}%
\thanks{Ahmed Naeem is with the School of Electrical Engineering and Computer Science, National University of Sciences and Technology, 44000, Islamabad, Pakistan (e-mail: ahmed.naeem@seecs.edu.pk).\\This work has been submitted to the IEEE for possible publication. Copyright may be transferred without notice, after which this version may no longer be accessible.}}
\markboth{Journal of \LaTeX\ Class Files,~Vol.~14, No.~8, July~2026}%
{Shell \MakeLowercase{\textit{et al.}}: Bare Demo of IEEEtran.cls for IEEE Journals}
\maketitle 
\begin{abstract}
Cell-free massive multiple-input multiple-output (CF-mMIMO) promises uniform service across the coverage area, and this service relies on accurate channel estimation for every user equipment (UE). These estimates are obtained from orthogonal pilot sequences, whose number is bounded by the channel coherence block. As a result, the sequences are reused as the UE density increases and the resulting pilot contamination limits the scalability. This paper proposes a joint user association (UA) and pilot assignment (PA) framework built on the adaptive phase-shifted (APS) pilot design, in which all UEs share a single full-band pilot and a per-UE phase shift applied in the frequency domain places their channel impulse responses (CIRs) in contiguous, non-overlapping intervals of the time domain at the access point (AP). The number of UEs that an AP can separate is thus set by the number of subcarriers rather than by the number of orthogonal sequences, and the UA is formulated as a $\mathbf{0/1}$ knapsack problem whose item weights are the numbers of taps and whose capacity is the number of subcarriers. Analysis and simulations show that the proposed design more than doubles the number of UEs scheduled by the conventional interleaved (CI) benchmark. It attains the same mean-squared error (MSE) as the CI benchmark under an equal total pilot power, and a lower MSE under an equal per-subcarrier power. It further improves the 5th-percentile connectivity and Jain's fairness index by penalizing the number of existing connections of a UE in the value assigned to each candidate link.
\end{abstract}
\section{Introduction}
\IEEEPARstart{A}{s} wireless networks evolve toward \ac{6G}, providing high and uniform data rates across the coverage area has become a pivotal objective~\cite{andrews2016we}. However, cell-based topologies cannot provide uniform \ac{QoS} and data rates across \acp{UE}, particularly for cell-edge \acp{UE}, and their coverage remains limited~\cite{11045549}. To address these limitations, the traditional cell-based topology gives way to the \ac{CF-mMIMO} paradigm~\cite{11373758}.

\par The canonical \ac{CF-mMIMO}, where every \ac{AP} serves every \ac{UE}, becomes impractical at scale because its fronthaul signaling and computational load increase drastically with the numbers of \acp{AP} and \acp{UE}~\cite{8761828}. In addition, the number of orthogonal pilots available to resolve the \ac{UE} channels at the \acp{AP} becomes insufficient~\cite{demir2021foundations}. To overcome the first of these limitations, user-centric \ac{CF-mMIMO} serves each \ac{UE} through a nearby subset of \acp{AP}~\cite{8000355}, which reduces this burden while preserving uniform coverage free of the cell-edge effect that is prominent in cell-based topology.

\par However, this reduction in fronthaul and processing load does not eliminate the second factor that limits the number of \acp{UE} and \acp{AP} a system can support. The limited interval allocated to \ac{UL} training within the coherence time restricts the number of mutually orthogonal pilot sequences that can be generated. As the \ac{UE} density increases, the number of \acp{UE} exceeds the number of available orthogonal sequences. At this point, reuse of the same sequence by multiple \acp{UE} across the network becomes unavoidable. An \ac{AP} that serves two pilot-sharing \acp{UE} then obtains a superposition of their channels, which is known as pilot contamination~\cite{10216314}. Consequently, the number of \acp{UE} served by an \ac{AP} without contamination is limited to the number of available orthogonal pilots~\cite{demir2021foundations}.
\par Under this limitation, the \ac{UA} and \ac{PA} decisions become critical. The \ac{UA} determines the serving \acp{AP} of each \ac{UE} and thus which \acp{UE} meet at a common \ac{AP}. The \ac{PA} assigns a pilot sequence to each \ac{UE} and thus determines which \acp{UE} share the same sequence. The channel estimation accuracy and the scalability are joint outcomes of the two decisions.
\subsection{Related Works}
\par In a scalable architecture, each \ac{UE} can be served by only a limited number of \acp{AP}, which has motivated extensive work on the \ac{UA}~\cite{9064545}. Early user-centric schemes fix the \ac{UA} by a heuristic rule in which each \ac{AP} serves a fixed number of the \acp{UE} with the strongest channels~\cite{8000355, buzzi2019user}. The \ac{UA} is later posed as an optimization problem. In~\cite{11083674}, the serving clusters are formed under fronthaul and per-\ac{AP} power limits, and a two-stage matched decision between the \acp{UE} and the \acp{AP} is adopted in~\cite{10015115}. An accelerated projected gradient method reduces the computational cost at scale~\cite{hao2023user}, and a deep Q-network placed at each \ac{UE} learns the decision~\cite{11369903}. In all the aforementioned \ac{UA} schemes, the pilot of a \ac{UE} is fixed before the clusters are formed. Another line of work takes the serving clusters as given and formulates the \ac{PA} as a conflict-avoidance problem through greedy updates~\cite{ngo2017cell}, graph coloring~\cite{9110802}, Hungarian matching~\cite{buzzi2021pilot}, and tabu search~\cite{liu2020tabu}. These approaches reduce pilot contamination. However, the size of the pilot pool remains unchanged. A third line of work decides the \ac{UA} and the \ac{PA} within a single procedure. A master \ac{AP} assigns the least-contaminated pilot~\cite{9064545}, and an interference-aware reward drives two iterative algorithms~\cite{chen2023improving}. In~\cite{9174860}, a structured access framework forms the clusters before the \ac{PA}, and an auction allocates the pilots after the \ac{UA} in~\cite{10091547}. A max-min distance rule performs the \ac{PA} before a greedy \ac{UA}~\cite{youn2024joint}. In contrast, a preference-based \ac{UA} is followed by a least-conflict \ac{PA} rule~\cite{alalwani2026scalable}. The two decisions remain sequential in these schemes, and the number of \acp{UE} that an \ac{AP} serves is limited by the number of orthogonal pilot sequences.

\par A separate line of work produces the pilots outside the code domain. In~\cite{jiang2021cellfree}, the pilot pool is enlarged with the frequency dimension: one resource unit of a resource block carries the pilot of a \ac{UE} while no other \ac{UE} transmits on it, and each \ac{UE} occupies one subcarrier of one symbol regardless of its channel. The subcarriers are partitioned into disjoint sets whose sizes follow from the data-rate requests of the \acp{UE} in~\cite{jiang2022opportunistic}, where each \ac{UE} is then served by a fixed number of \acp{AP} with the strongest \acp{LSFC}. The rule that maps the requests to the sets remains unspecified. Superimposing the pilot on the data extends the pool to the length of the coherence interval, at the cost of an interference term between the two and a split of the transmit power~\cite{zhang2021superimposed}. A precoded variant reduces this interference at the cost of a reduced number of data symbols~\cite{garg2022generalized}. Pilot hopping identifies a \ac{UE} by a sequence that spans several coherence intervals~\cite{becirovic2022activity}. The number of distinguishable \acp{UE} thereby scales with the number of intervals, whereas two \acp{UE} may still collide within one interval.
\par Despite substantial research, joint \ac{UA} and \ac{PA} strategies that scale with the \ac{UE} density by generating the pilots outside the code domain remain underdeveloped in the literature. In the schemes that focus on the \ac{UA} and the \ac{PA}, the pilots are in every case selected from a fixed pool of orthogonal sequences. On the other hand, the schemes that generate the pilots enlarge this pool, yet the enlarged resource is allocated independently of the \ac{UA} and the \ac{PA}.

\subsection{Motivation and Key Contributions}
\par The motivation of this work is to increase the number of \acp{UE} that a user-centric \ac{CF-mMIMO} system serves while the system remains scalable. \ac{UE}-specific phase shifts applied in the frequency domain make the \acp{UE} separable in the time domain at the \ac{AP}. The key contributions are as follows:
\begin{itemize}
\item A joint \ac{UA} and \ac{PA} framework is proposed for user-centric \ac{CF-mMIMO} on top of an \ac{APS} pilot design. The \ac{UA} is formulated as a $0/1$ knapsack problem, which maximizes the total value of the selected items under a capacity constraint. The \ac{PA} is carried out by the \ac{APS} pilot design, in which the phase shift of a selected \ac{UE} follows from its position in the knapsack solution. By placing the \acp{CIR} of the served \acp{UE} in contiguous, non-overlapping delay-domain intervals, the proposed design substantially increases the number of \acp{UE} that an \ac{AP} can serve.

\item The knapsack profit increases with a link's received power and decreases with its number of taps, favoring links that consume less \ac{AP} capacity. A fairness penalty, whose strength is set by a tunable exponent, grows with the number of \acp{AP} already serving a \ac{UE} and directs the remaining capacity toward the least-connected \acp{UE}. Each \ac{AP}'s knapsack is solved exactly by dynamic programming in pseudo-polynomial time, at a cost bilinear in the numbers of \acp{AP} and \acp{UE}.

\item The proposed framework is evaluated through Monte Carlo simulations across varying \ac{AP} and \ac{UE} densities, cyclic prefix lengths, and fairness penalty exponents. Under these settings, an \ac{AP} serves at least twice the number of \acp{UE} served by the \ac{CI} benchmark, while the \ac{MSE} is not degraded under an equal total pilot power and is lower under an equal per-subcarrier power. The control signaling overhead and the computational complexity of the proposed algorithm are also quantified.

\end{itemize}
\subsection{Notation}
\par Italic letters denote scalars (e.g., $x$). Bold uppercase letters denote matrices (e.g., $\mathbf{X}$). Calligraphic letters denote sets (e.g., $\mathcal{X}$). The set of $a \times b$ complex matrices is denoted by $\mathbb{C}^{a \times b}$ and the set of $a \times b$ binary matrices is denoted by $\{0,1\}^{a \times b}$. The operator $\mathbb{E}[\cdot]$ denotes the expectation. Accents distinguish related variables, where $(\hat{\cdot})$ denotes an estimate, $(\tilde{\cdot})$ denotes a restricted set, $(\bar{\cdot})$ denotes a phase-shifted variable, and $(\cdot)'$ denotes the matrix or the set obtained after the proposed \ac{UA}. For a set $\mathcal{X}$, the notation $|\mathcal{X}|$ denotes its cardinality. For a scalar $x$, the notation $|x|$ denotes its magnitude. The notations $\mathcal{CN}(\mu,\sigma^{2})$ and $\mathcal{N}(\mu,\sigma^{2})$ denote the circularly symmetric complex and the real Gaussian distributions with mean $\mu$ and variance $\sigma^{2}$, respectively.

\section{System Model}
\begin{figure}[!t]
    \centering
    \includegraphics[width=0.9\linewidth, trim={1.75cm 2.9cm 3cm 1.5cm}, clip]{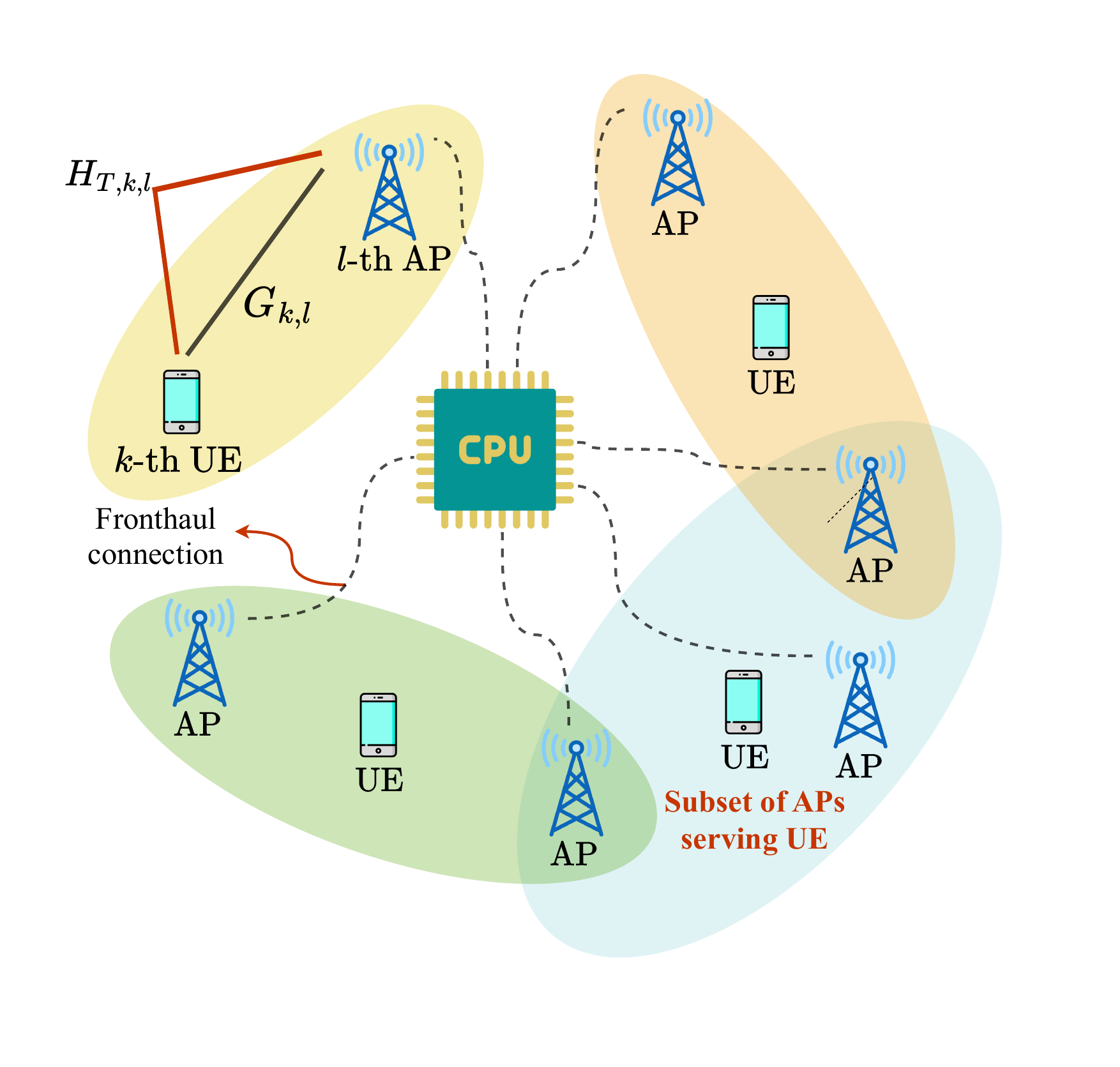}
   \caption{The considered \ac{CF-mMIMO} architecture, where distributed \acp{AP} connected to a \ac{CPU} via fronthaul links serve the \acp{UE}. Each ellipse encloses the user-centric subset of \acp{AP} serving a \ac{UE}.}  
   \label{fig:cf_model}
\end{figure}

We consider a \ac{CF-mMIMO} system with $L$ \acp{AP} and $K$ \acp{UE} that are independently and uniformly distributed over the coverage area. Fig.~\ref{fig:cf_model} illustrates the \ac{CF-mMIMO} architecture considered in this paper together with the user-centric subset of \acp{AP} serving each \ac{UE}. All the \acp{AP} are connected to the \ac{CPU} through fronthaul links, and the system operates in \acf{TDD} mode~\cite{9064545}. Both the \acp{AP} and the \acp{UE} are equipped with a single antenna. The phase shift is applied at each \ac{UE}, hence the separation of the \acp{CIR} does not depend on the number of antennas at the \ac{AP}, and the single-antenna case is considered.

\subsection{Channel Model}
\par The large-scale fading is modeled by a log-distance path loss model with a free-space reference given by the Friis equation~\cite{rappaport2002wireless}. Let $\xi_{k,l}^{\mathrm{dB}}$ denote the \ac{LSFC} between the $k$-th \ac{UE} and the $l$-th \ac{AP}. These coefficients are assumed to be known at the \ac{CPU}. The reference path loss $\mathrm{PL}_{\mathrm{dB}}(d_0)$ at the reference distance $d_0$ is given by
\begin{equation}
\mathrm{PL}_{\mathrm{dB}}(d_0) = 20 \log_{10}(d_0) + 20 \log_{10}(f) - 147.55,
\label{eq:pathloss_reference}
\end{equation}
\begin{equation}
\xi_{k,l}^{\mathrm{dB}}
=
\mathrm{PL}_{\mathrm{dB}}(d_0)
+
10\alpha
\log_{10}\!\left(\frac{d_{k,l}}{d_0}\right)
+ Z_{k,l},
\label{eq:large_scale_fading}
\end{equation}
where $d_{k,l}$ is the distance in meters between the $k$-th \ac{UE} and the $l$-th \ac{AP}, $f$ is the carrier frequency in Hz, $\alpha$ is the path loss exponent, and $Z_{k,l} \sim \mathcal{N}(0, \sigma_{\mathrm{sh}}^{2})$ models the shadow fading in dB and is independent across the \ac{UE}-\ac{AP} links.
\par The \ac{UL} channel between the $k$-th \ac{UE} and the $l$-th \ac{AP} is modeled as a frequency-selective, time-invariant multipath channel with $G_{k,l}$ resolvable taps~\cite{4267831}. The baseband-equivalent \ac{CIR} is expressed as
\begin{equation}
H_{T,k,l}(\tau) = \sum_{g=0}^{G_{k,l}-1} h_{T,k,l}(g)\, \delta\!\left(\tau - \tau_{k,l}(g)\right),
\label{eq:cir}
\end{equation}
where $h_{T,k,l}(g) \sim \mathcal{CN}(0, 1/G_{k,l})$ and $\tau_{k,l}(g)$ denote the complex channel gain and the propagation delay of the $g$-th path between the $k$-th \ac{UE} and the $l$-th \ac{AP}, respectively. The gains are independent across the paths, so that the small-scale channel has unit power and the large-scale attenuation is carried by the \ac{LSFC}. The channel is assumed to be quasi-static over a block of multiple \ac{OFDM} symbols, while varying over longer time intervals. The corresponding \ac{CFR} is given by
\begin{equation}
H_{F,k,l}(\nu,s) = \sum_{g=0}^{G_{k,l}-1} h_{T,k,l}(g)\, e^{-j2\pi \nu \Delta f \tau_{k,l}(g)},
\label{eq:cfr}
\end{equation}
where $\Delta f$ is the subcarrier spacing. The number of resolvable taps $G_{k,l}$ is determined by the delay spread of the multipath channel, which varies slowly and can be estimated in advance~\cite{8007240}. Accordingly, $G_{k,l}$ is assumed to be known at each \ac{AP}, and this assumption does not require the tracking of the instantaneous channel~\cite{4267831}.

\subsection{Large-Scale Fading Coefficient-Based Masking Process}  \label{2B}
\par The candidate links of each \ac{AP} are determined from the \acp{LSFC} by a masking process. The power received from the $k$-th \ac{UE} at the $l$-th \ac{AP} is given by
\begin{equation}
P_{k,l}^{\mathrm{dBm}} = P_{\mathrm{t}}^{\mathrm{dBm}} - \xi_{k,l}^{\mathrm{dB}},
\label{eq:received_power}
\end{equation}
where $P_{\mathrm{t}}^{\mathrm{dBm}}$ is the \ac{UL} transmit power of a \ac{UE} and is identical for all \acp{UE}. To limit the fronthaul overhead, the links that exceed a predefined threshold $\gamma_{\mathrm{th}}^{\mathrm{dBm}}$ are collected in the matrix $\mathbf{S} \in \{0,1\}^{K \times L}$. The element $S_{k,l}$ indicates whether the $k$-th \ac{UE} is eligible to be served by the $l$-th \ac{AP} and is given by
\begin{equation}
\label{eq:first_stage_LFSC_matrix}
S_{k,l} =
\begin{cases}
1, & \text{if } P_{k,l}^{\mathrm{dBm}} \geq \gamma_{\mathrm{th}}^{\mathrm{dBm}},\\
0, & \text{otherwise}.
\end{cases}
\end{equation}
\par The candidate set of \acp{AP} for the $k$-th \ac{UE} is defined as $\mathcal{M}_k = \{\, l \in \{1,\ldots,L\} \mid S_{k,l} = 1 \,\}$, and the candidate set of \acp{UE} for the $l$-th \ac{AP} is defined as $\mathcal{D}_l = \{\, k \in \{1,\ldots,K\} \mid S_{k,l} = 1 \,\}$. The proposed $0/1$ knapsack-based joint \ac{UA} and \ac{PA} algorithm then yields the \ac{UA} matrix $\mathbf{S}' \in \{0,1\}^{K \times L}$. From this matrix, the serving clusters are defined as $\mathcal{M}'_k = \{\, l \in \{1,\ldots,L\} \mid S'_{k,l} = 1 \,\}$ for the $k$-th \ac{UE} and $\mathcal{D}'_l = \{\, k \in \{1,\ldots,K\} \mid S'_{k,l} = 1 \,\}$ for the $l$-th \ac{AP}. 
\subsection{Adaptive Phase-Shifted Pilot Design for UL Training}
\par This subsection presents the \ac{APS} pilot design adopted for the \ac{UL} training phase~\cite{sumer2025adaptive}.

\subsubsection{Transmitter Side Process}
\par Let $N$ denote the number of subcarriers and $N_{\mathrm{sym}}$ the number of \ac{OFDM} symbols, indexed by $\nu \in \{0,\dots,N-1\}$ and $s \in \{0,\dots,N_{\mathrm{sym}}-1\}$, respectively. Let $X_{F}(\nu,s)$ denote the base pilot sequence shared by all \acp{UE}, which has unit modulus, $|X_{F}(\nu,s)| = 1$. Each \ac{AP} estimates the channels of its served \acp{UE} on a dedicated \ac{OFDM} symbol, on which the \acp{UE} in $\mathcal{D}'_l$ multiply the base sequence by their own phase shifts. A \ac{UE} that is served by several \acp{AP} applies the phase shift assigned by each serving \ac{AP}, and transmits on a separate \ac{OFDM} symbol of the \ac{UL} training phase for each of them.\footnote{The same requirement applies to the \ac{CI} benchmark, since a \ac{UE} served by several \acp{AP} is in general assigned a different interleaved subcarrier set at each of them. Both schemes employ the same number of \ac{UL} training symbols.} The intra-\ac{AP} pilot contamination is eliminated by construction, while the inter-\ac{AP} separation relies on \ac{AP}-orthogonal training symbols. The transmitted sequence of the $k$-th \ac{UE} at the $l$-th \ac{AP} is denoted as
\begin{equation}
\bar{X}_{F,k,l}(\nu,s) = X_{F}(\nu,s)\, \phi_{k,l}(\nu),
\label{eq:tx_seq}
\end{equation}
where $\phi_{k,l}(\nu)$ is the phase shift of the $k$-th \ac{UE} at the $l$-th \ac{AP}. The \ac{UA} is completed before the training phase, and the serving clusters are known. The phase shift applied to the $\nu$-th subcarrier is given by
\begin{equation}
\phi_{k,l}(\nu) = e^{-j \frac{2\pi \nu\, n_{k,l}}{N}},
\label{eq:phi_kl}
\end{equation}
where $n_{k,l}$ is the cumulative delay offset of the $k$-th \ac{UE}, obtained by summing the numbers of resolvable taps of the preceding \acp{UE}
\begin{equation}
n_{k,l} = \sum_{q=1}^{r-1} G_{\pi_l(q),\, l},
\label{eq:n_kl}
\end{equation}
where $\pi_l$ denotes the ordering of the serving cluster $\mathcal{D}'_l$, $\pi_l(q)$ is the global index of the \ac{UE} that occupies the $q$-th position of this ordering, $r$ is the position of the $k$-th \ac{UE} with $k = \pi_l(r)$, and $G_{\pi_l(q),\,l}$ is the number of resolvable taps between the \ac{UE} at position $q$ and the $l$-th \ac{AP}.
\par The offset of a \ac{UE} follows from its position in the ordering, and the \ac{PA} is thereby obtained from the \ac{UA} rather than from a separate optimization. Since the frequency-domain phase shift corresponds to a cyclic shift in the time domain, the \acp{CIR} remain contiguous and non-overlapping only as long as the total number of their taps does not exceed the $N$ samples of the time domain. This condition is defined for the $l$-th \ac{AP} as
\begin{equation}
\label{eq:APS_capacity}
\sum_{k \in \mathcal{D}'_l} G_{k,l} \leq N.
\end{equation}
\par The time-domain \ac{OFDM} signal transmitted by the $k$-th \ac{UE} at the $n$-th sample of the $s$-th \ac{OFDM} symbol is obtained by applying the \acf{IFFT} to $\bar{X}_{F,k,l}(\nu,s)$
\begin{equation}
\bar{X}_{T,k,l}(n,s)
=
\frac{1}{\sqrt{N}}
\sum_{\nu=0}^{N-1}
\bar{X}_{F,k,l}(\nu,s)
\, e^{j\frac{2\pi \nu n}{N}}.
\end{equation}
\subsubsection{Receiver Side Process}
\par A \ac{CP} of $N_{\mathrm{cp}}$ samples is appended to each \ac{OFDM} symbol before transmission. The transmitted pilots propagate through the frequency-selective channel and are processed by the FFT at the receiver. The received frequency-domain signal at the $l$-th \ac{AP} is the superposition of the phase-shifted pilots of the served \acp{UE} given as
\begin{equation}
Y_{F,l}(\nu,s)
=
\sum_{k \in \mathcal{D}_l'}
H_{F,k,l}(\nu,s)\,
\bar{X}_{F,k,l}(\nu,s)
+ W_{F,l}(\nu,s),
\end{equation}
where $W_{F,l}(\nu,s) \sim \mathcal{CN}(0,\sigma_{w}^{2})$ is the \ac{AWGN} sample and $\sigma_{w}^{2}$ denotes the noise variance.
\begin{figure*}[t]
\centering
\begin{subfigure}[b]{0.49\textwidth}
    \centering
    \includegraphics[
        width=\linewidth,
        trim={1cm 1cm 1cm 0.8cm},
        clip
    ]{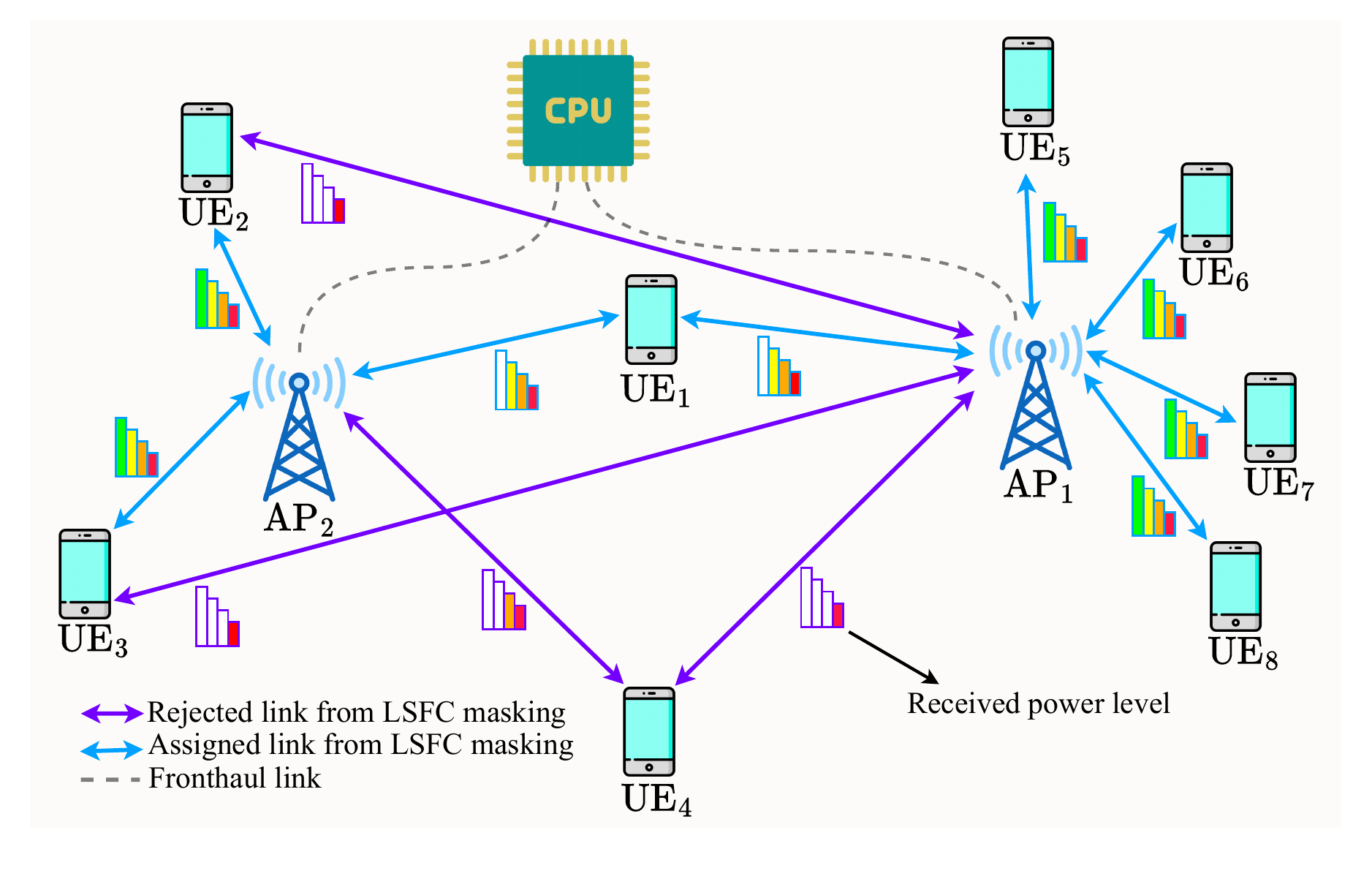}
    \caption{}
    \label{fig:lsfc_a}
\end{subfigure}
\hfill
\begin{subfigure}[b]{0.49\textwidth}
    \centering
    \includegraphics[
        width=\linewidth,
        trim={1cm 0.7cm 1.7cm 0.6cm},
        clip
    ]{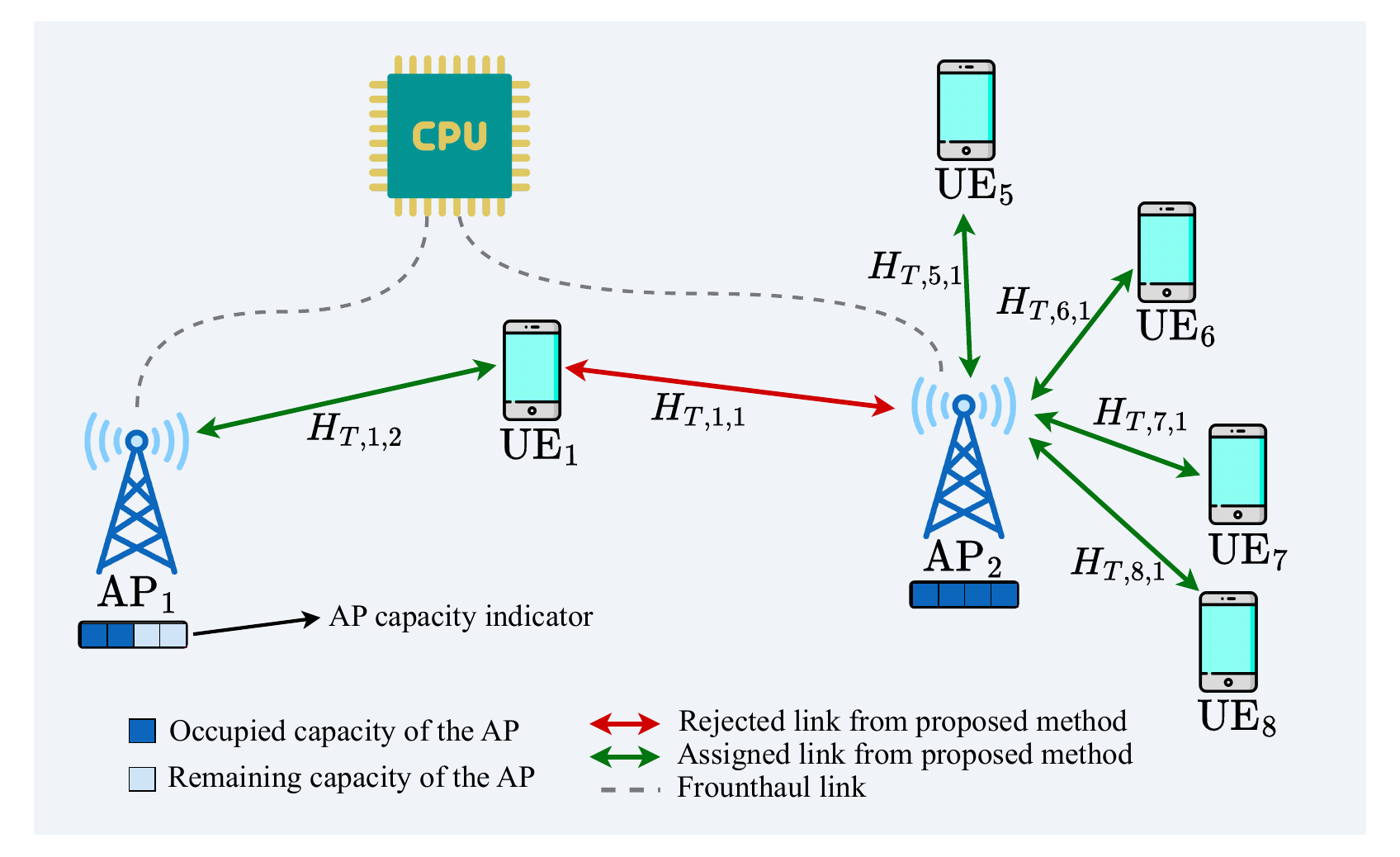}
    \caption{}
    \label{fig:lsfc_b}
\end{subfigure}
\caption{Illustration of the proposed joint \ac{UA} and \ac{PA} framework. (a)~The \ac{LSFC}-based masking stage, where the bar on each link shows the received power $P_{k,l}^{\mathrm{dBm}}$ relative to the threshold $\gamma_{\mathrm{th}}^{\mathrm{dBm}}$. (b)~The knapsack-based \ac{UA} stage, where the bar at each \ac{AP} shows its occupied capacity.}
\label{fig:lsfc_framework}
\end{figure*}
\par The received pilot signal is then divided by the base pilot sequence to estimate the channels of the served \acp{UE}, and the resulting estimate contains the superposition of their \acp{CFR}
\begin{equation}
\hat{H}_{F,l}(\nu, s) = \frac{Y_{F,l}(\nu, s)}{X_F(\nu, s)}.
\end{equation}
An $N$-point \ac{IFFT} of $\hat{H}_{F,l}(\nu,s)$ yields the time-domain estimate $\hat{H}_{T,l}(n,s)$ at the $l$-th \ac{AP}, in which the \acp{UE} are separated
\begin{equation}
\hat{H}_{T,l}(n,s) =
\frac{1}{\sqrt{N}}
\sum_{\nu=0}^{N-1}
\hat{H}_{F,l}(\nu,s)\,
e^{j\frac{2\pi \nu n}{N}}.
\end{equation}

\subsubsection{The \texorpdfstring{\ac{CIR}}{CIR} Separation Process}
\par The \acp{CIR} of the served \acp{UE} occupy contiguous and non-overlapping intervals of $\hat{H}_{T,l}(n,s)$, as determined by the phase shifts. The $l$-th \ac{AP} extracts the contribution of the $k$-th \ac{UE} and reconstructs its \ac{CFR} $\hat{H}_{F,k,l}(\nu,s)$ by an FFT
\begin{equation}
\hat{H}_{F,k,l}(\nu,s)
=
\sqrt{N}
\sum_{n=n_{k,l}}^{n_{k,l}+G_{k,l}-1}
\hat{H}_{T,l}(n,s)\,
e^{-j\frac{2\pi \nu n}{N}}.
\label{eq:HF_est}
\end{equation}
\section{Proposed Joint User Association and Pilot Assignment}
\label{sec:proposed_user_association}
\par This section presents the proposed joint \ac{UA} and \ac{PA} algorithm. The constraint in~\eqref{eq:APS_capacity} allows an \ac{AP} to serve only a subset of its candidate \acp{UE}. This subset is the serving cluster $\mathcal{D}'_l$, and it is obtained from a $0/1$ knapsack problem. This problem maximizes the total profit of the selected items subject to a capacity constraint, and each item is either selected or excluded~\cite{martello1990knapsack}. Each candidate \ac{UE} of an \ac{AP} is an item, the weight of an item is its number of resolvable taps $G_{k,l}$, and the capacity is $N$.

\par Fig.~\ref{fig:lsfc_framework} illustrates the two stages of the proposed \ac{UA} framework. In the first stage, shown in Fig.~\ref{fig:lsfc_a}, the path loss between each \ac{UE} and each \ac{AP} is computed from~\eqref{eq:pathloss_reference} and~\eqref{eq:large_scale_fading}, and the links whose received power exceeds the predefined threshold $\gamma_{\mathrm{th}}^{\mathrm{dBm}}$ are retained as candidates~\eqref{eq:first_stage_LFSC_matrix}. A \ac{UE} may be a candidate for several \acp{AP}. For example, $\mathrm{UE}_1$ appears in the candidate sets of both $\mathrm{AP}_1$ and $\mathrm{AP}_2$ because its received power exceeds the threshold at both. In the second stage, shown in Fig.~\ref{fig:lsfc_b}, the knapsack determines which candidates are selected subject to the capacity constraint in~\eqref{eq:APS_capacity}. In this example, $\mathrm{UE}_5$--$\mathrm{UE}_8$ attain the highest profit at $\mathrm{AP}_1$, and their $G_{5,1}$--$G_{8,1}$ occupy the capacity of $\mathrm{AP}_1$. The profit of $\mathrm{UE}_1$ is the lowest among the candidates at $\mathrm{AP}_1$, hence the link between $\mathrm{UE}_1$ and $\mathrm{AP}_1$ is rejected. $\mathrm{AP}_2$ has free capacity, hence the link between $\mathrm{UE}_1$ and $\mathrm{AP}_2$ is selected.
\subsection{Fairness-Aware Profit Formulation}
\label{subsec:profit}
\par In the proposed joint \ac{UA} and \ac{PA} algorithm, a knapsack problem is solved for each \ac{AP}, because the capacity constraint in~\eqref{eq:APS_capacity} is defined per \ac{AP}. After each \ac{AP} is processed, the numbers of connections of the \acp{UE} are updated. Let $\mathcal{A}_{\text{proc}}$ denote the set of \acp{AP} that have already been processed. The number of connections of the $k$-th \ac{UE} is defined as
\begin{equation}
\label{eq:kanpsacak_subsubset}
d_k = \sum_{m \in \mathcal{A}_{\text{proc}}} S'_{k,m}, \quad \forall k \in \{1,\dots,K\}.
\end{equation}
\par The weight $w_{k,l}$ and the capacity $C$ are modeled as
\begin{equation}
\label{eq:Knapsacak_weight_and_capacity}
    w_{k,l} = G_{k,l}, \qquad C = N,
\end{equation}
where both follow from the non-overlap condition in~\eqref{eq:APS_capacity}.
\par At the $l$-th \ac{AP}, only the \acp{UE} of $\mathcal{D}_l$ whose weight is at least one and at most $C$ remain eligible. Therefore, the feasible set is
\begin{equation}
\tilde{\mathcal{D}}_l = \left\{ k \in \mathcal{D}_l \mid 1 \leq w_{k,l} \leq C\right\}.
\label{eq:feasible_set}
\end{equation}
\par Once $\tilde{\mathcal{D}}_l$ is fixed, a profit is computed for each link in $\tilde{\mathcal{D}}_l$, and these profits define the objective of the knapsack. The profit increases with the received power of the link and decreases with the number of taps that the link occupies. To provide fairness, it also decreases with the number of connections $d_k$ that the \ac{UE} has already accumulated. A \ac{UE} that has not been assigned any connection by the preceding \acp{AP} enters the knapsack at its full profit. This mechanism protects the connectivity of the least-connected \acp{UE}. Then, the profit $p_{k,l}$ is defined as
\begin{equation}
p_{k,l} = \frac{P_{k,l}^{\mathrm{lin}}}{G_{k,l}\,(1+d_k)^{\beta}},
\label{eq:priority_metric}
\end{equation}
where $P_{k,l}^{\mathrm{lin}}$ is the received power of the link on a linear scale and $\beta$ is the fairness penalty exponent, which governs the trade-off between the average connectivity and the \ac{UE} fairness.
\par Because the profit in~\eqref{eq:priority_metric} is inversely proportional to $G_{k,l}$, a \ac{UE} with strong received power and few taps attains a higher profit than one with the same power and more taps. Such a \ac{UE} occupies a small part of the capacity, and the remaining capacity accommodates further \acp{UE}. The number of connections $d_k$ balances this preference. It reduces the profit of a \ac{UE} as the number of its connections increases and gives priority to the \acp{UE} with $d_k=0$. The profit scales with the received power of the link, while the penalty depends only on the number of connections of the \ac{UE}. Consequently, each additional connection yields a smaller marginal profit for that \ac{UE}. With the profits and the weights defined, the knapsack problem of the $l$-th \ac{AP} is formulated as
\begin{equation}
\label{eq:knapsack}
\begin{aligned}
\max_{\{S'_{k,l}\}_{k\in\tilde{\mathcal{D}}_l}}\! \!\!\quad & \sum_{k\in\tilde{\mathcal{D}}_l} p_{k,l}\, S'_{k,l} \\
\text{s.t.} \!\!\!\quad & \sum_{k\in\tilde{\mathcal{D}}_l} w_{k,l}\, S'_{k,l} \le C, \!\!\quad S'_{k,l}\in\{0,1\},\ \forall\, k\in\tilde{\mathcal{D}}_l,
\end{aligned}
\end{equation}
where the $l$-th column of the \ac{UA} matrix $\mathbf{S}'$ collects the decisions of the $l$-th \ac{AP}.
\begin{algorithm}[t!]
\caption{0/1 Knapsack-Based Joint \ac{UA} and \ac{PA}}
\label{alg:proposed_system}
\begin{algorithmic}[1]
\STATE \textbf{Input:} $K, L, N, P_{\mathrm{t}}^{\mathrm{dBm}}, \xi_{k,l}^{\mathrm{dB}}, P_{k,l}^{\mathrm{lin}}, G_{k,l}, \beta, \gamma_{\mathrm{th}}^{\mathrm{dBm}}$
\STATE \textbf{Output:} $\mathbf{S}' \in \{0,1\}^{K \times L}$, $\{\mathcal{D}'_l\}$, $\{n_{k,l}\}$
\STATE \textbf{Initialize:} $\mathbf{S}' \leftarrow \mathbf{0}_{K \times L}, \ d_k \leftarrow 0 \ \forall k, \ C \leftarrow N$
\STATE \textbf{Stage 1: \ac{LSFC}-based masking}
\STATE $\mathcal{D}_l \leftarrow \{\, k \in \{1,\dots,K\} \mid P_{k,l}^{\mathrm{dBm}} \ge \gamma_{\mathrm{th}}^{\mathrm{dBm}} \,\}, \ \forall\, l$ \hfill \eqref{eq:received_power}, \eqref{eq:first_stage_LFSC_matrix}
\STATE \textbf{Stage 2: Fairness-Aware 0/1 Knapsack \ac{UA} and \ac{PA}}
\STATE $w_{k,l} \leftarrow G_{k,l}, \ \forall\, k \in \mathcal{D}_l, \ \forall\, l$ \hfill \eqref{eq:Knapsacak_weight_and_capacity}
\FOR{$l = 1, \dots, L$}
    \STATE $\tilde{\mathcal{D}}_l \leftarrow \{k \in \mathcal{D}_l \mid 1 \le w_{k,l} \le C\}, \ v_l \leftarrow |\tilde{\mathcal{D}}_l|$ \hfill \eqref{eq:feasible_set}
    \STATE Index $\tilde{\mathcal{D}}_l$ by $i \in \{1,\dots,v_l\}$, with $k(i)$ the global index of candidate $i$
    \STATE \textbf{Profit assignment:} \hfill \eqref{eq:priority_metric}
    \FOR{$i = 1, \dots, v_l$}
        \STATE $p_{k(i),l} \leftarrow P_{k(i),l}^{\mathrm{lin}}/\bigl(G_{k(i),l}\,(1+d_{k(i)})^{\beta}\bigr)$
    \ENDFOR
    \STATE \textbf{Dynamic programming:} $V \leftarrow \mathbf{0}$, $B \leftarrow \mathbf{0}$ \hfill \eqref{eq:dp_recursion}, \eqref{eq:knapsack_decision}
    \FOR{$i = 1, \dots, v_l$}
        \FOR{$c = C$ \textbf{down to} $w_{k(i),l}$}
            \IF{$V[c - w_{k(i),l}] + p_{k(i),l} > V[c]$}
                \STATE $V[c] \leftarrow V[c - w_{k(i),l}] + p_{k(i),l}, \ B[i, c] \leftarrow 1$
            \ENDIF
        \ENDFOR
    \ENDFOR
    \STATE \textbf{Backtracking and update:} $c \leftarrow C$ \hfill \eqref{eq:backtracking_in}, \eqref{eq:kanpcask_upgrate}
    \FOR{$i = v_l$ \textbf{down to} $1$}
        \STATE $S'_{k(i),l} \leftarrow B[i, c]$
        \IF{$S'_{k(i),l} = 1$}
            \STATE $d_{k(i)} \leftarrow d_{k(i)} + 1, \ c \leftarrow c - w_{k(i),l}$ \hfill \eqref{eq:kanpsacak_subsubset}
        \ENDIF
    \ENDFOR
    \STATE $\mathcal{D}'_l \leftarrow \{\, k \in \tilde{\mathcal{D}}_l \mid S'_{k,l} = 1 \,\}$
    \STATE \textbf{Pilot assignment:} $\eta \leftarrow 0$ \hfill \eqref{eq:n_kl}
    \FOR{$i = 1, \dots, v_l$}
        \IF{$S'_{k(i),l} = 1$}
            \STATE $n_{k(i),l} \leftarrow \eta, \ \eta \leftarrow \eta + w_{k(i),l}$
        \ENDIF
    \ENDFOR
\ENDFOR
\end{algorithmic}
\end{algorithm}

\subsection{Dynamic Programming Solution}
\label{subsec:dp}
\par The problem in~\eqref{eq:knapsack} is NP-hard, but its capacity is bounded by $N$, and it is solved by dynamic programming based on the Bellman equation~\cite{martello1990knapsack}, which builds the optimal selection of the $l$-th \ac{AP} from the optimal selections of smaller subproblems. Let $v_l = |\tilde{\mathcal{D}}_l|$ denote the number of feasible candidates at the $l$-th \ac{AP}. These candidates are processed one at a time and are indexed by a local position $i \in \{1,\dots,v_l\}$. The \ac{UA} matrix is indexed by the global \ac{UE} index $k$, and $k(i)$ denotes the global index of the candidate at local position $i$. In this local indexing, the profit and the weight of a candidate are $p_{k(i),l}$ and $w_{k(i),l}$. Let $c \in \{0,1,\dots,C\}$ denote the remaining capacity. The value table $V(i,c)$ is defined as the maximum total profit that the first $i$ candidates can provide within a capacity of $c$. Each entry is obtained from the entries of the preceding candidate by the recursion
\begin{equation}
V(i,c) \! =\!
\left\{
\begin{array}{@{}l@{}}
V(i-1,c), \\
\quad \text{if } w_{k(i),l} > c,\\[3pt]
\max\bigl\{ V(i-1,c),\, V(i-1,c-w_{k(i),l}) + p_{k(i),l} \bigr\}, \\
\quad \text{if } w_{k(i),l} \le c.
\end{array}\right.
\label{eq:dp_recursion}
\end{equation}
The base case of the recursion is $V(0,c) = 0$ for all $c$. The profit is zero before any candidate is selected. Each candidate is selected at most once.
\par The recursion in~\eqref{eq:dp_recursion} gives the maximum profit, but it does not record which candidates attain it. The binary decision table $B(i,c)$ is filled in the same forward pass and marks the states at which selecting the $i$-th candidate yields a higher profit than rejecting it
\begin{equation}
B(i,c) \!=\!
\begin{cases}
1, & \text{if } w_{k(i),l} \le c \ \& \\
   & \ \ \!\!V(i-1,c-w_{k(i),l}) + p_{k(i),l} > V(i-1,c), \\[3pt]
0, & \text{otherwise}.
\end{cases}
\label{eq:knapsack_decision}
\end{equation}
\par The selection is obtained by backtracking through $\mathbf{B}$ from the state $(v_l, C)$. Let $c_i$ denote the remaining capacity at step $i$, with $c_{v_l} = C$. The decision of the $i$-th candidate is given by $B(i,c_i)$, and the \ac{UA} matrix is updated as
\begin{equation}
\label{eq:backtracking_in}
S'_{k(i),l} = B(i, c_i),
\end{equation}
 where $B(i,c_i)$ is the optimal decision of~\eqref{eq:knapsack} for $k=k(i)$. The remaining capacity is updated as
\begin{equation}
\label{eq:kanpcask_upgrate}
c_{i-1} = c_i - B(i, c_i)\, w_{k(i),l},
\end{equation}
where the capacity decreases by the weight $w_{k(i),l}$ for a selected candidate and remains unchanged for a rejected candidate. All the steps described above are summarized in Algorithm~\ref{alg:proposed_system}.

\subsection{Computational Complexity Analysis}
\par The computational cost of the proposed algorithm is quantified against the \ac{CI} benchmark, and its scaling with the numbers of \acp{AP} and \acp{UE} is assessed. The complexity is measured by the number of arithmetic and comparison operations that are executed when the serving clusters are updated. The \acp{LSFC} and the numbers of resolvable taps are assumed to be known. Their acquisition is common to both methods and is excluded from the comparison. The \ac{LSFC}-based masking step is applied identically by both methods and is excluded for the same reason.
\par In the \ac{CI} scheme, each \ac{AP} sorts its candidates by received power and selects one candidate for each of its disjoint subcarrier sets. The sorting requires $\mathcal{O}(|\mathcal{D}_l|\log|\mathcal{D}_l|)$ operations per \ac{AP}~\cite{cormen2022introduction}, and the total complexity of the \ac{CI} benchmark is given by
\begin{equation}
\label{eq:complexity_ci}
\mathcal{C}_{\mathrm{CI}} = \mathcal{O}\Bigl(\sum_{l=1}^{L}|\mathcal{D}_l|\log|\mathcal{D}_l|\Bigr).
\end{equation}
\par In the proposed algorithm, computing the profit in~\eqref{eq:priority_metric} for every candidate of an \ac{AP} requires $\mathcal{O}(|\mathcal{D}_l|)$ operations. The knapsack in~\eqref{eq:knapsack} is solved by the recursion in~\eqref{eq:dp_recursion}, and one entry of the value table $V(i,c)$ is computed for each pair of a candidate and a capacity value. The capacity is $N$, and the value table requires $\mathcal{O}(|\tilde{\mathcal{D}}_l|\,N)$ operations per \ac{AP}~\cite{martello1990knapsack}. The backtracking adds $\mathcal{O}(|\tilde{\mathcal{D}}_l|)$ operations, and the update of the number of connections in~\eqref{eq:kanpsacak_subsubset} requires $\mathcal{O}(1)$ operations per selected candidate. The value table is larger than these terms by the factor $N$, and it determines the cost at an \ac{AP}. The total complexity becomes
\begin{equation}
\label{eq:complexity_prop}
\mathcal{C}_{\mathrm{APS}} = \mathcal{O}\Bigl(N\sum_{l=1}^{L}|\tilde{\mathcal{D}}_l|\Bigr).
\end{equation}
\par The two complexity expressions differ in the factor that multiplies the number of candidates of an \ac{AP}. This factor is $\log|\mathcal{D}_l|$ for the \ac{CI} benchmark and $N$ for the proposed algorithm. The factor $N$ is the additional cost of the knapsack. Since $N$ is a fixed system parameter bounded by the coherence bandwidth and does not scale with $L$ or $K$, the complexity in~\eqref{eq:complexity_prop} is bilinear in both $L$ and $K$ for fixed $N$.

\subsection{Control Signaling Analysis}
\par The \ac{CIR} separation in~\eqref{eq:HF_est} requires that each \ac{UE} applies the offset assigned to it during the \ac{UA}. The \ac{CPU} computes these offsets according to~\eqref{eq:n_kl} and forwards them to the \acp{AP} over the fronthaul links. Each \ac{AP} then transmits the offset to each served \ac{UE} on a downlink control channel, similar to the interleaved subcarrier offset signaling adopted in current standards~\cite{3gpp38211}. This transmission is the control signaling overhead of the proposed scheme. A \ac{UE} applies a separate offset at every serving \ac{AP}. Each offset is one of $N$ possible values and requires $\log_2 N$ control bits per link. The \ac{CI} scheme assigns a different subset of the subcarriers to each \ac{UE} that it serves. Let $\rho_{\mathrm{p}}$ denote the pilot ratio, that is, the fraction of the subcarriers that a \ac{UE} occupies, and let $U^{\mathrm{CI}} = 1/\rho_{\mathrm{p}}$ denote the number of \acp{UE} that an \ac{AP} serves. Each subset index is one of $1/\rho_{\mathrm{p}}$ possible values and requires $\log_2 (1/\rho_{\mathrm{p}})$ control bits per link. The average numbers of control bits received by a \ac{UE} under the proposed scheme and under the \ac{CI} benchmark are denoted by $Q^{\mathrm{APS}}$ and $Q^{\mathrm{CI}}$, respectively, and are given by
\begin{equation}
\label{eq:control_overhead}
Q^{\mathrm{APS}} = \frac{\log_2 N}{K}\sum_{l=1}^{L} |\mathcal{D}'_l|, \qquad
\!\!\!\!\!\!\!Q^{\mathrm{CI}} = \frac{L\,U^{\mathrm{CI}}}{K}\log_2\!\left(\frac{1}{\rho_{\mathrm{p}}}\right).
\end{equation}
These offsets change only when the serving clusters are updated. The signaling is repeated once per cluster update, and its cost stays negligible relative to the payload transmitted between two consecutive cluster updates.\footnote{The proposed scheme uses the numbers of resolvable taps $G_{k,l}$ to place the \acp{CIR} in non-overlapping intervals. These taps are estimated at the \ac{AP} and are assumed to be available there in both schemes, since both apply the same time-domain noise suppression. Thus, the \ac{MSE} comparison is made on equal terms. This estimation is common to the two schemes and is excluded from the comparison, which accounts only for the signaling of the offsets. The taps themselves are not signaled to the \acp{UE} in either scheme. However, in the proposed scheme, the offsets $n_{k,l}$ that are computed from them in~\eqref{eq:n_kl} are signaled to the \acp{UE}, which apply them as phase shifts. The procedures for estimating $G_{k,l}$ are described in~\cite{4267831, rappaport2015wideband}.}
\section{Simulation Results}
\par The proposed algorithm is evaluated for different \ac{AP} and \ac{UE} densities through four metrics, namely the average connectivity, the $5$th-percentile connectivity, Jain's fairness index, and the \ac{MSE}. The average connectivity is denoted as
\begin{equation}
d^{\mathrm{avg}} = \frac{1}{K}\sum_{k=1}^{K}\sum_{l=1}^{L} S'_{k,l}.
\label{eq:avg_connectivity}
\end{equation}
The $5$th-percentile connectivity is the value below which the numbers of connections of $5\%$ of the \acp{UE} fall, and it characterizes the least-connected \acp{UE}.
\begin{table}[t]
\renewcommand{\arraystretch}{1.3}
\caption{Simulation Parameters and Network Configurations}
\label{tab:simulation_parameters}
\centering
\begin{tabular}{ll}
\hline
\textbf{Parameter} & \textbf{Value} \\ \hline
Service area size & $100 \times 100\text{ m}^2$ \\
Carrier frequency ($f$) & $2.4\text{ GHz}$ \\
Transmit power ($P_{\mathrm{t}}^{\mathrm{dBm}}$) & $20\text{ dBm}$ \\
Path loss exponent ($\alpha$) & $3$ \\
Reference distance ($d_0$) & $1\text{ m}$ \\
Shadow fading standard deviation ($\sigma_{\mathrm{sh}}$) & $3\text{ dB}$ \\
LSFC-based masking threshold ($\gamma_{\mathrm{th}}^{\mathrm{dBm}}$) & $-60\text{ dBm}$ \\
Number of subcarriers ($N$) & $64$ \\
CP length ($N_{\mathrm{cp}}$) & $8, 16, 32$ \\
Pilot ratio ($\rho_{\mathrm{p}}$) & $1/8, 1/4, 1/2$ \\
Fairness penalty exponent ($\beta$) & $0, 5, 10, 15$ \\
\hline
\end{tabular}
\end{table}
\par In user-centric \ac{CF-mMIMO}, the \ac{UA} produces serving clusters of different sizes~\cite{chen2023improving}. For this reason, the average number of connections alone does not capture the uniformity of these connections across the \acp{UE}. This uniformity is measured by Jain's fairness index~\cite{jain1984fairness}, which is computed over the number of \ac{AP} connections of each \ac{UE}, following the connectivity-based formulation of~\cite{9946428}. From the final \ac{UA} matrix $\mathbf{S}'$ obtained in Section~\ref{sec:proposed_user_association}, Jain's fairness index is denoted as
\begin{equation}
J = 
\frac{\left(\sum_{k=1}^{K} \sum_{l=1}^{L} S'_{k,l}\right)^2}
{K \sum_{k=1}^{K} \left(\sum_{l=1}^{L} S'_{k,l}\right)^2},
\label{eq:jains_fairness}
\end{equation}
where $J \in [1/K, 1]$. The value $J = 1$ corresponds to perfect fairness, in which every \ac{UE} is served by the same number of \acp{AP}. Smaller values indicate that most of the connections are held by a subset of the \acp{UE}.
\par The quality of the channel estimates obtained over these connections is evaluated separately by the \ac{MSE}, which is computed only over the served links. For each link, the \ac{MSE} between $\hat{H}_{F,k,l}$ in~\eqref{eq:HF_est} and $H_{F,k,l}$ in~\eqref{eq:cfr} is denoted as
\begin{equation}
\mathrm{MSE}_{k,l} \!= \!\frac{1}{N N_{\mathrm{sym}}}\sum_{s=0}^{N_{\mathrm{sym}}-1}\sum_{\nu=0}^{N-1} \left| \hat{H}_{F,k,l}(\nu,s) - H_{F,k,l}(\nu,s) \right|^{2}.
\label{eq:mse}
\end{equation}
The average over the served links is
\begin{equation}
\mathrm{MSE}^{\mathrm{avg}} = \frac{1}{\sum_{l=1}^{L}|\mathcal{D}'_l|}
\sum_{l=1}^{L} \sum_{k \in \mathcal{D}'_l} \mathrm{MSE}_{k,l}.
\label{eq:mse_avg}
\end{equation}
The \ac{SNR} is defined as $1/\sigma_{w}^{2}$, where $\sigma_{w}^{2}$ is the noise variance at every \ac{AP}.
\subsection{MSE Performance}
\begin{figure}[!t]
\centering
\begin{subfigure}{\columnwidth}
    \centering
    \includegraphics[width=0.9\linewidth]{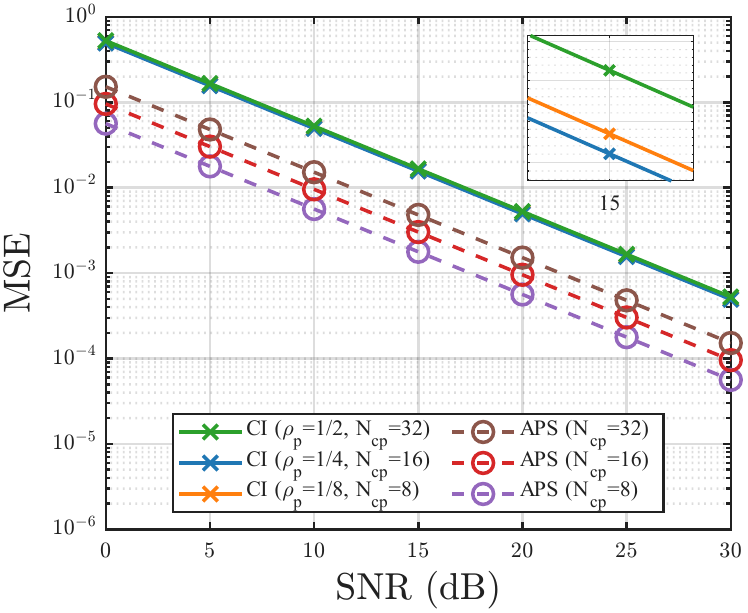}
    \caption{}
    \label{fig:mse_a}
\end{subfigure}
\begin{subfigure}{\columnwidth}
    \centering
    \includegraphics[width=0.9\linewidth]{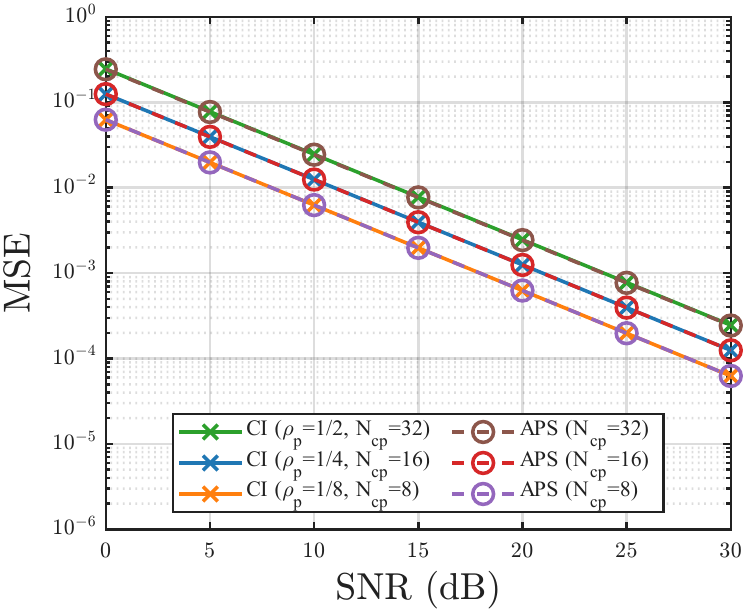}
    \caption{}
    \label{fig:mse_b}
\end{subfigure}
\caption{\ac{UL} training \ac{MSE} for the \ac{APS} design with $N_{\mathrm{cp}} \in \{8, 16, 32\}$ and the \ac{CI} benchmark with $\rho_{\mathrm{p}} \in \{1/8, 1/4, 1/2\}$: (a)~per-subcarrier power constraint and (b)~total power constraint.}
\label{fig:mse_performance}
\end{figure}
\par The following \ac{CP} ratios $\rho_{\mathrm{cp}} \in \{1/8,\, 1/4,\, 1/2\}$ are considered, and the channel of every \ac{UE} is generated with at most $N_{\mathrm{cp}}-1$ resolvable taps. Therefore, reliable channel estimation in the \ac{CI} benchmark requires at least $N_{\mathrm{cp}}$ pilot subcarriers per \ac{UE}~\cite{4267831}, and each \ac{UE} is assigned $N_{\mathrm{cp}}$ of the $N$ subcarriers, $\rho_{\mathrm{cp}} = N_{\mathrm{cp}}/N = \rho_{\mathrm{p}}$. The number of \acp{UE} scheduled by an \ac{AP} in the \ac{CI} benchmark is limited to
$U^{\mathrm{CI}} = N/N_{\mathrm{cp}} = 1/\rho_{\mathrm{p}}$. Unlike the \ac{CI} benchmark, the \ac{APS} pilot design allocates all subcarriers to every \ac{UE}.
\begin{figure*}[!t]
\centering
\begin{subfigure}{0.40\textwidth}
    \centering
    \includegraphics[width=\linewidth]{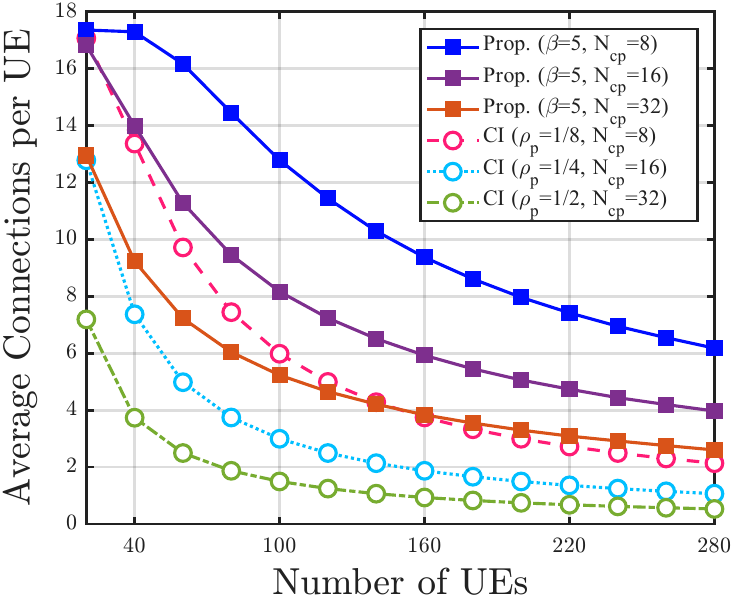}
    \caption{}
    \label{fig:avg_cp}
\end{subfigure}
\hspace{0.05\textwidth}
\begin{subfigure}{0.40\textwidth}
    \centering
    \includegraphics[width=\linewidth]{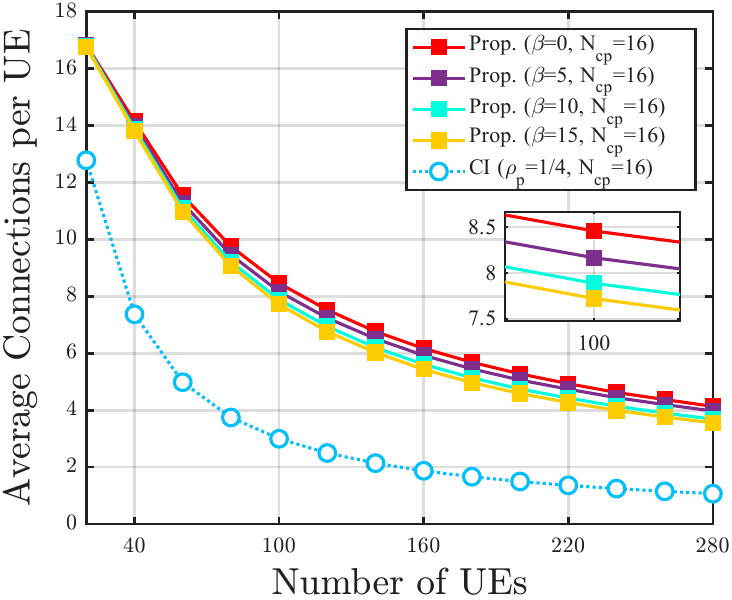}
    \caption{}
    \label{fig:avg_beta}
\end{subfigure}
\begin{subfigure}{0.40\textwidth}
    \centering
    \includegraphics[width=\linewidth]{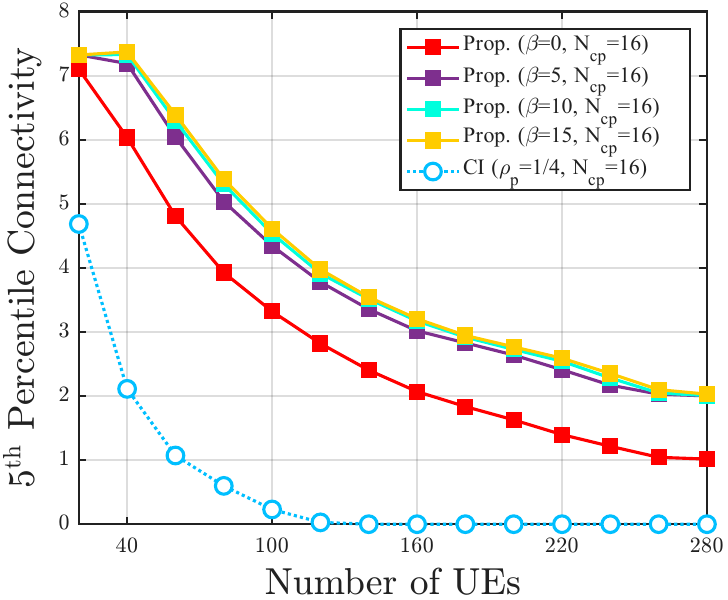}
    \caption{}
    \label{fig:cell_beta}
\end{subfigure}
\hspace{0.05\textwidth}
\begin{subfigure}{0.40\textwidth}
    \centering
    \includegraphics[width=\linewidth]{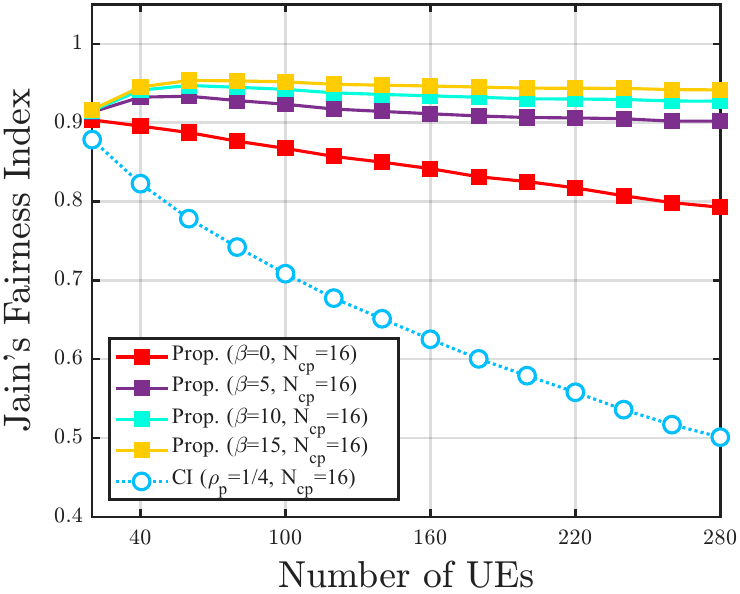}
    \caption{}
    \label{fig:fair_beta}
\end{subfigure}
\caption{Performance evaluation versus the number of \acp{UE} with the number of \acp{AP} fixed at $75$. (a)~Average connectivity for different $N_{\mathrm{cp}}$ at $\beta=5$, (b)~Average connectivity for different $\beta$, (c)~$5$th-percentile connectivity for different $\beta$, and (d)~Jain's fairness index for different $\beta$.}
\label{fig:performance_results}
\end{figure*}
\par Fig.~\ref{fig:mse_performance} presents the \ac{UL} training \ac{MSE} performance for the parameters in Table~\ref{tab:simulation_parameters} under two power constraints~\cite{sumer2025adaptive}. Under the per-subcarrier power constraint in Fig.~\ref{fig:mse_a}, every design transmits at a fixed power per active subcarrier. Under the total power constraint in Fig.~\ref{fig:mse_b}, the per-subcarrier power of the \ac{CI} benchmark is boosted to equalize the total transmitted power of the two designs. The \acp{UE} are free of interference in both designs, hence the estimation error results only from the \ac{AWGN}. In particular, the \ac{CIR} separation in~\eqref{eq:HF_est} takes the FFT over the $G_{k,l}$ samples of the interval of a \ac{UE} rather than over all $N$ samples. Only the noise within this interval contributes to $\hat{H}_{F,k,l}$, and the retained noise power scales with $G_{k,l}$. A larger $N_{\mathrm{cp}}$ produces a larger $G_{k,l}$, which increases the number of noise-accumulating taps. Therefore, the \ac{MSE} of both designs increases with $N_{\mathrm{cp}}$. In Fig.~\ref{fig:mse_a}, a \ac{UE} transmits its pilot on all $N$ subcarriers in the \ac{APS} design, whereas it transmits on only $\rho_{\mathrm{p}} N$ subcarriers in the \ac{CI} benchmark. The \ac{CI} benchmark delivers $\rho_{\mathrm{p}}$ times the pilot energy of the \ac{APS} design to the \ac{AP} and attains a larger \ac{MSE}. In Fig.~\ref{fig:mse_b}, the \ac{CI} benchmark compensates for this loss by boosting its per-subcarrier power, and the two designs attain the same \ac{MSE}. Consequently, the gap is set by the total power allocated to the pilot signal rather than by the estimation capability of the two designs~\cite{1193803}.
\begin{figure*}[!t]
\centering
\begin{subfigure}{0.40\textwidth}
    \centering
    \includegraphics[width=\linewidth]{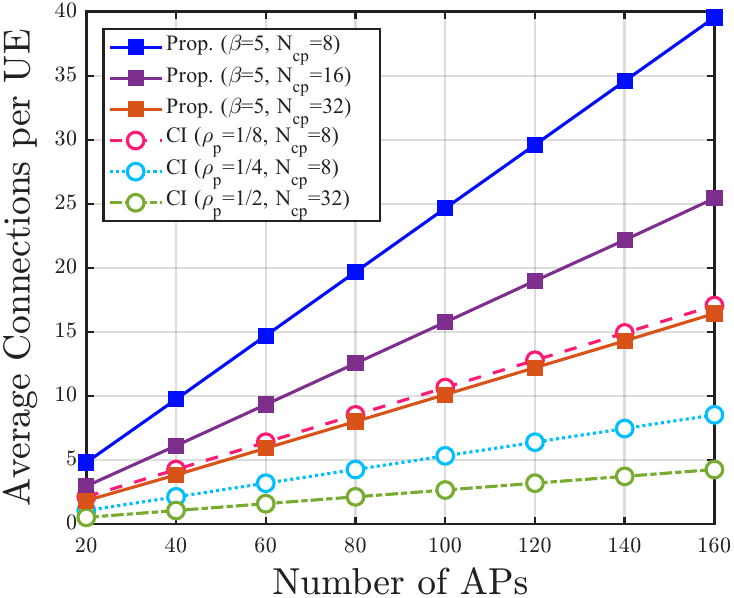}
    \caption{}
    \label{fig:avg_cp_ap}
\end{subfigure}
\hspace{0.05\textwidth}
\begin{subfigure}{0.40\textwidth}
    \centering
    \includegraphics[width=\linewidth]{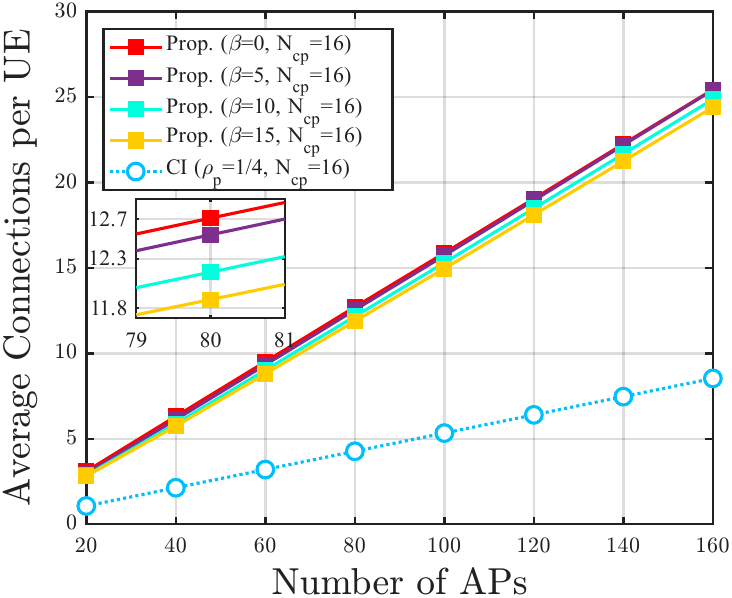}
    \caption{}
    \label{fig:avg_beta_ap}
\end{subfigure}
\begin{subfigure}{0.40\textwidth}
    \centering
    \includegraphics[width=\linewidth]{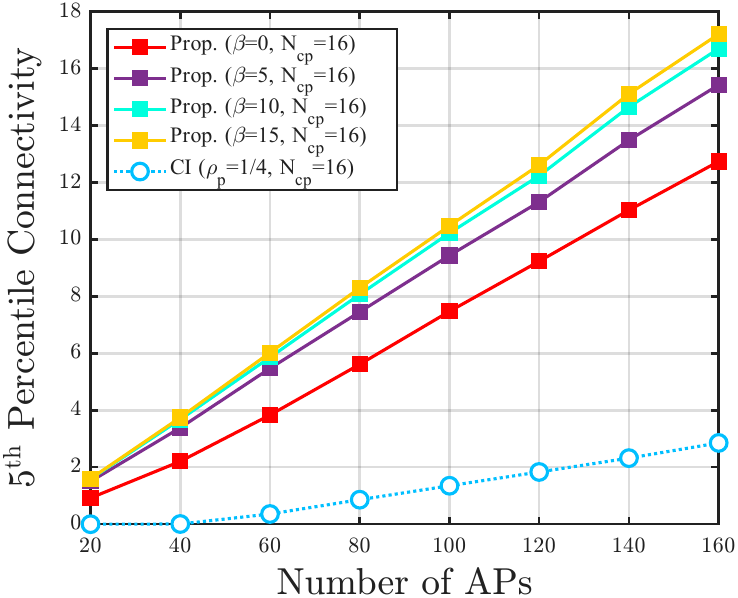}
    \caption{}
    \label{fig:cell_beta_ap}
\end{subfigure}
\hspace{0.05\textwidth}
\begin{subfigure}{0.40\textwidth}
    \centering
    \includegraphics[width=\linewidth]{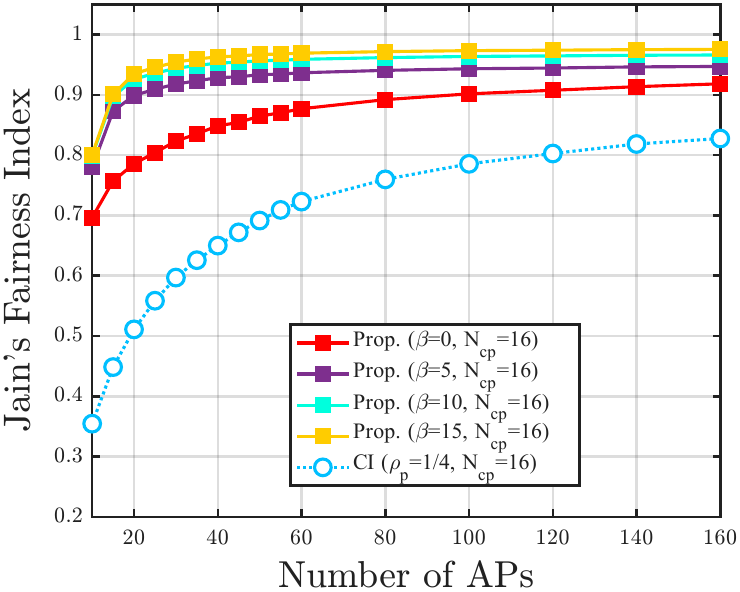}
    \caption{}
    \label{fig:fair_beta_ap}
\end{subfigure}
\caption{Performance evaluation versus the number of \acp{AP} with the number of \acp{UE} fixed at $75$. (a)~Average connectivity for different $N_{\mathrm{cp}}$ at $\beta=5$, (b)~Average connectivity for different $\beta$, (c)~$5$th-percentile connectivity for different $\beta$, and (d)~Jain's fairness index for different $\beta$.}
\label{fig:performance_results_ap}
\end{figure*}
\subsection{Joint User Association and Pilot Assignment Performance}
\par The average number of scheduled \acp{UE} of the proposed joint \ac{UA} and \ac{PA}
algorithm is first expressed and then evaluated through simulations. Each $G_{k,l}$ is modeled by a normal distribution truncated to $[1,\,N_{\mathrm{cp}}-1]$, which leaves at least one sample of the \ac{CP} as a guard margin~\cite{tse2005fundamentals}. Under this distribution, the expected number of resolvable taps is denoted as
\begin{equation}
\mathbb{E}[G_{k,l}] = \frac{N_{\mathrm{cp}}}{2}.
\label{eq:expected_taps}
\end{equation}
Accordingly, the average number of scheduled \acp{UE} of the \ac{APS} design at an \ac{AP} satisfies
\begin{equation}
\mathbb{E}[U_l^{\mathrm{APS}}] \geq \frac{N}{\mathbb{E}[G_{k,l}]} = \frac{2N}{N_{\mathrm{cp}}}, \quad \forall l \in \{1,\dots,L\},
\label{eq:expected_aps}
\end{equation}
where $U_l^{\mathrm{APS}}$ denotes the number of \acp{UE} scheduled by the \ac{APS} design
at the $l$-th \ac{AP}. The ratio of this number to that of the \ac{CI} benchmark is
\begin{equation}
\frac{\mathbb{E}[U_l^{\mathrm{APS}}]}{U^{\mathrm{CI}}} \geq \frac{2N/N_{\mathrm{cp}}}{1/\rho_{\mathrm{p}}} = 2.
\label{eq:aps_ratios}
\end{equation}
\subsubsection{Effect of UE Density}
\label{subsubsec:ue_density}
\par Fig.~\ref{fig:performance_results} presents the results obtained when the number of \acp{UE} is varied and the number of \acp{AP} is fixed at $75$. The simulation parameters are listed in Table~\ref{tab:simulation_parameters}.

\par In Fig.~\ref{fig:avg_cp}, the proposed method increases the number of scheduled \acp{UE}
substantially relative to the \ac{CI} benchmark, and the size of this increase is governed
mainly by $N_{\mathrm{cp}}$. For $100$ \acp{UE} and the same $N_{\mathrm{cp}}$, the ratio of the average connectivity of the proposed method to that of the \ac{CI} benchmark is $2.13$ for $N_{\mathrm{cp}}=8$, $2.72$ for $N_{\mathrm{cp}}=16$, and $3.48$ for $N_{\mathrm{cp}}=32$. All three values exceed the ratio in~\eqref{eq:aps_ratios}, and the ratio increases with $N_{\mathrm{cp}}$. The reason is that an \ac{AP} serves $1/\rho_{\mathrm{p}}$ \acp{UE} in the \ac{CI} benchmark, and this number is halved when $\rho_{\mathrm{p}}$ is doubled. In contrast, the number of \acp{UE} that an \ac{AP} serves in the proposed method is set by the mean number of taps of the selected \acp{UE} as in~\eqref{eq:expected_aps}. As $N_{\mathrm{cp}}$ increases, the range of $G_{k,l}$ widens and its expected value in~\eqref{eq:expected_taps} grows. The candidate set of an \ac{AP} is determined by the threshold in~\eqref{eq:first_stage_LFSC_matrix} and is independent of $N_{\mathrm{cp}}$. Accordingly, the \ac{AP} selects fewer \acp{UE} from the same pool, and the links with few taps and strong received power gain a higher priority. As a result, the mean number of taps of the selected \acp{UE} stays below the expected value in~\eqref{eq:expected_taps}. This gap between the expected value and the mean of the selected \acp{UE} widens with $N_{\mathrm{cp}}$, and the ratio increases accordingly.
\par Furthermore, for $160$ \acp{UE}, the corresponding ratios between the proposed method and the \ac{CI} benchmark reach $2.50$, $3.15$, and $4.12$, which shows that the ratio increases with the number of \acp{UE}. The reason is that a larger candidate set contains more \acp{UE} with few taps and strong received power, and these \acp{UE} gain an even larger selection priority as $N_{\mathrm{cp}}$ increases. Consequently, the proposed method performs substantially better than the \ac{CI} benchmark both under channel conditions with a large delay spread and under dense \ac{UE} deployments.

\par As can be seen from Fig.~\ref{fig:avg_beta}, the average connectivity per \ac{UE} decreases as $\beta$ increases. The reason is that the profit of a \ac{UE} with several connections is divided by $(1+d_k)^{\beta}$ in~\eqref{eq:priority_metric}, and a larger $\beta$ strengthens this reduction. The selection priority of the \acp{UE} that already hold many connections therefore decreases, and the \acp{UE} with few connections are prioritized. This effect is observed in Fig.~\ref{fig:cell_beta}, where the number of connections of the least-connected \acp{UE} increases substantially with $\beta$ relative to the fairness-unaware case $\beta=0$. The two figures together expose the trade-off between the total number of connections in the network and the connectivity of the least-connected \acp{UE}, and $\beta$ is the parameter that adjusts this trade-off.

\par The uniformity of the connections across the \acp{UE} is evaluated in Fig.~\ref{fig:fair_beta} through Jain's fairness index in~\eqref{eq:jains_fairness}. The fairness index of the \ac{CI} benchmark declines rapidly as the number of \acp{UE} increases. Each \ac{AP} serves only the $U^{\mathrm{CI}}$ \acp{UE} with the highest received power, hence the same \acp{UE} are selected at most of the \acp{AP} and the variance of the number of connections across the \acp{UE} grows. In contrast, the index of the proposed method increases with $\beta$, which is consistent with Fig.~\ref{fig:cell_beta}. Moreover, the index remains nearly constant up to $300$ \acp{UE} for $\beta \geq 10$. Therefore, the fairness penalty preserves the distribution of the connections even under a heavy load, whereas without the penalty, that is, for $\beta=0$, the index decreases as the number of \acp{UE} increases.

\subsubsection{Effect of AP Density}
\par Fig.~\ref{fig:performance_results_ap} presents the results obtained when the number of \acp{AP} is varied and the number of \acp{UE} is fixed at $75$.
\par As can be seen from Fig.~\ref{fig:avg_cp_ap}, the average connectivity of both methods increases linearly with the number of \acp{AP}. For $60$ \acp{AP}, the ratio of the average connectivity of the proposed method to that of the \ac{CI} benchmark is $2.29$ for $N_{\mathrm{cp}}=8$, $2.92$ for $N_{\mathrm{cp}}=16$, and $3.69$ for $N_{\mathrm{cp}}=32$. Since the curves of both methods are linear, this ratio remains nearly constant over the examined numbers of \acp{AP}.
\begin{table*}[!t]
\renewcommand{\arraystretch}{1.2}
\caption{Control signaling overhead and computational complexity for three values of $K$ at $L=75$ \acp{AP}, with $N=64$.}
\label{tab:overhead_complexity}
\centering
\begin{tabular}{llccccccc}
\hline
\multirow{2}{*}{$K$} & \multirow{2}{*}{\textbf{Method}} &
\multicolumn{3}{c}{\textbf{Control bits per link}} &
\multicolumn{3}{c}{\textbf{Control bits per \ac{UE}}} &
\multirow{2}{*}{\textbf{Complexity} $\mathcal{C}$ $(\times 10^{4})$} \\\cline{3-5}\cline{6-8}
 & & $\rho_{\mathrm{p}}=1/8$ & $1/4$ & $1/2$ & $\rho_{\mathrm{p}}=1/8$ & $1/4$ & $1/2$ & \\
\hline
\multirow{3}{*}{$50$}
 & \ac{CI} benchmark     & 3.0 & 2.0 & 1.0 & 34.0  & 11.9 & 3.0  & 0.316 \\
 & Proposed, $\beta=0$   & 6.0 & 6.0 & 6.0 & 101.1 & 76.3 & 49.6 & 5.565 \\
 & Proposed, $\beta=10$  & 6.0 & 6.0 & 6.0 & 100.8 & 74.2 & 47.3 & 5.565 \\
\hline
\multirow{3}{*}{$75$}
 & \ac{CI} benchmark     & 3.0 & 2.0 & 1.0 & 23.8 & 8.0  & 2.0  & 0.548 \\
 & Proposed, $\beta=0$   & 6.0 & 6.0 & 6.0 & 90.1 & 60.9 & 38.9 & 8.346 \\
 & Proposed, $\beta=10$  & 6.0 & 6.0 & 6.0 & 88.6 & 57.9 & 36.4 & 8.346 \\
\hline
\multirow{3}{*}{$100$}
 & \ac{CI} benchmark     & 3.0 & 2.0 & 1.0 & 18.0 & 6.0  & 1.5  & 0.800 \\
 & Proposed, $\beta=0$   & 6.0 & 6.0 & 6.0 & 78.4 & 50.8 & 32.4 & 11.113 \\
 & Proposed, $\beta=10$  & 6.0 & 6.0 & 6.0 & 75.8 & 47.5 & 30.0 & 11.113 \\
\hline
\end{tabular}
\end{table*}
\par In Fig.~\ref{fig:avg_beta_ap}, the curves obtained for different $\beta$ almost coincide as the number of \acp{AP} increases. A larger number of \acp{AP} provides more knapsack capacity per \ac{UE}, and fewer candidates are rejected, so the total number of connections becomes largely independent of $\beta$. The few remaining rejections do not change the average connectivity appreciably, but the \acp{UE} in whose favor they are made determine the lower tail of the distribution.

\par In Fig.~\ref{fig:cell_beta_ap}, a clear difference is observed between $\beta=0$ and $\beta=5$, whereas the curves for $\beta=10$ and $\beta=15$ are almost on top of each other. The reason is that the profit is divided by $(1+d_k)^{\beta}$ in~\eqref{eq:priority_metric}. As $\beta$ increases, the small differences in the number of connections become more dominant in the profit. As a result, the \acp{UE} with fewer connections are prioritized more strongly. This prioritization strengthens rapidly for small values of $\beta$, whereas once the penalty is strong enough, a further increase of $\beta$ changes the selection progressively less. Thus, the curves move closer to each other as $\beta$ increases.

\par Fig.~\ref{fig:fair_beta_ap} shows that the improvement for the least-connected \acp{UE} is also reflected in the distribution of the connections over the network. Consistently with Fig.~\ref{fig:cell_beta_ap}, the fairness index increases with $\beta$ and the curves move closer to each other as $\beta$ grows. The index saturates toward the maximum fairness value of $1$ as the number of \acp{AP} increases and the gap to the \ac{CI} benchmark narrows, although the \ac{CI} benchmark remains below the proposed method over the entire \ac{AP} range. Consequently, the loss in average connectivity caused by the fairness penalty becomes negligible as the \ac{AP} density increases. In addition, the gains in the $5$th-percentile connectivity and in the fairness index are preserved. Overall, under a dense \ac{AP} deployment, fairness is obtained without a loss in average connectivity.
\subsection{Control Signaling and Complexity Trade-off}
\par Table~\ref{tab:overhead_complexity} reports the control signaling overhead and the computational complexity. The control signaling per link is determined by the number of possible index values, and the proposed method requires $\log_2 N = 6$ bits while the \ac{CI} benchmark requires $\log_2(1/\rho_{\mathrm{p}})$ bits. The control signaling per \ac{UE} is $78.4$ bits for the proposed method and $18.0$ bits for the benchmark at $K=100$ and $\rho_{\mathrm{cp}}=1/8$. Although the difference per link is only a factor of two, the larger gap follows from the substantially larger number of links that the proposed method establishes. In addition, the control signaling per \ac{UE} decreases as the number of \acp{UE} grows, and it decreases further at a larger pilot ratio and at a larger $\beta$. This decrease occurs because the time-domain resource of the network is fixed and is shared among more \acp{UE}. As a result, the number of connections per \ac{UE} decreases, together with the data transmitted to each \ac{UE}.
\par The measured complexity of the proposed method is more than ten times that of the \ac{CI} benchmark for all three values of $K$. Moreover, it remains $0.111\times10^{4}$ per \ac{UE}, which verifies the linearity in the number of \acp{UE} stated in~\eqref{eq:complexity_prop}. The proposed method requires a higher control signaling overhead and a higher computational complexity than the \ac{CI} benchmark. On the other hand, it provides more than twice the number of scheduled \acp{UE} at the same channel estimation \ac{MSE}, a higher connectivity for the least-connected \acp{UE}, and a fairer distribution of the connections over the network.

\section{Conclusion}
\par This paper presented a scalable joint \ac{UA} and \ac{PA} framework for user-centric \ac{CF-mMIMO}. The \ac{APS} pilot design was adopted, and the joint \ac{UA} and \ac{PA} of multiple \acp{UE} to multiple \acp{AP} was formulated as a single $0/1$ knapsack problem. Within the same formulation, a fairness-aware penalty in the profit shifts the remaining capacity of an \ac{AP} toward the least-connected \acp{UE}. Although the formulated problem is NP-hard, its capacity is bounded by the number of subcarriers. Therefore, the knapsack of each \ac{AP} is solved by dynamic programming in pseudo-polynomial time, at a cost that is bilinear in the numbers of \acp{AP} and \acp{UE}. The resulting framework attains more than twice the number of \acp{UE} scheduled by the
\ac{CI} benchmark and improves both the $5$th-percentile connectivity and Jain's fairness index. Furthermore, it attains the same \ac{MSE} as the benchmark under an equal total pilot power and a lower \ac{MSE} under an equal per-subcarrier power. These gains cost a few signaling bits per link and additional computational complexity. Both are incurred only when the serving clusters are updated, hence their cost remains negligible. Future work will evaluate the uplink and downlink spectral efficiency under the resulting channel estimates and will address the pilot contamination that arises when the proposed framework is combined with multi-user spatial multiplexing.


\end{document}